\documentclass[openacc]{rsproca_new}

\begin{document}

\title{Animal Synchrony and agents’ segregation}

\author{
Laura P. Schaposnik$^{1,2,3}$,  Sheryl Hsu$^{4}$, and Robin I.M. Dunbar$^{2}$}

\address{ $^{1}$University of Illinois at Chicago,   USA. \\
   $^{2}$ University of Oxford, UK.\\
    $^{3}$  Mathematical Science Research Institute, USA.\\
   $^{4}$ Stanford University, USA.  }

\subject{}


\corres{Laura P. Schaposnik\\
\email{schapos@uic.edu}}

\begin{abstract}
In recent years it has become evident the need of understanding how failure of coordination imposes constraints on the size of stable groups that highly social mammals can live in. We examine here the forces that keep animals together as a herd and others that drive them apart. Different phenotypes (e.g. genders) have different rates of gut fill, causing them to spend different amounts of time performing activities. By modeling a group as a set of semi-coupled oscillators on a disc, we show that the members of the group may become less and less coupled until the group dissolves and breaks apart. We show that when social bonding creates a stickiness, or gravitational pull, between pairs of individuals, fragmentation is reduced. 
\end{abstract}


\begin{fmtext}
%
%
%
%
%
%
%
%
%
%

\section{Introduction}
The formation of groups amongst different animals has long been studied both from the mathematical perspective as well as from a behavioural perspective (e.g. see \cite{[1]} and \cite{[2]}). In particular, for many mammal species, outside of the mating season males and females tend to form their own groups leading to  sexual segregation. Although it had previously been thought that this  was due to habitat segregation as males and females needed different types of habitat, it has been shown that this is not the sole cause \cite{[2]}. In many cases, groups fragment because the animals’ activity schedules  become  desynchronised \cite{[3]}-\cite{[D]}. This is especially likely to happen when differences in body size result in some animals wanting to go to rest when others still need to continue feeding. This can happen even when animals collectively agree on a direction of motion and a time to move in that direction (e.g., see \cite{[4], more1,more2,more3}).

\end{fmtext}


\maketitle

  In recent years, there have been increasing collaborations between mathematicians and biologists in the study of animals and their behaviors. Mathematicians have been able to develop models using data collected by biologists and test these models as hypotheses. These have led to many innovations such as applying bifurcation theory to population biology \cite{[6]},  coordination during travel \cite{travel}, or entropy analysis to model the optimal size and structure of social networks \cite{[E],[F]}. Of particular relevance to the present case has been mathematical analyses of behavioural coordination in group-living animals. Conradt \cite{[5]}, for example, created an index to measure the amount of sexual segregation in animals and used data collected by other scientists to show that sexual segregation in red deer and feral soay sheep was not simply caused by different genders having a preference for different types of habitat \cite{[2]}. The decision making processes involved have been modeled as an agent-based coupled model \cite{[1]}. In such settings, the members of the group decide on a direction of travel and whilst each agent has a preferred angle, they are gradually influenced by the position of other agents. As time passes, the agents move along a circle to the angle they desire, and at the end of the process, all the agents either agree on one angle or congregate in multiple groups that want to move in different directions, in which case the group breaks up into subgroups.

Most of these approaches have focussed on how animals coordinate movements when traveling in groups. However, an important converse problem is why and when groups fragment under these circumstances. This issue is important because it may impose limits on the size of group that animals can live in stable social groups of the kind found in primates and many other mammals (notably equids, camelids, delphinids and elephantids)  \cite{[2],[3]}. Understanding how failure of coordination imposes constraints on the size of stable groups thus goes beyond the kind of joiner/leaver models \cite{Kra} that have dominated classic behavioural ecology theory.  In these kinds of groups, bonded relationships between individuals create complex networks that create a form of gravitational drag that holds individuals in defined orbits with respect to each other \cite{[Dunbar]}. In the present paper, we expand on the work in \cite{[1]} by creating a model which allows for both   different activities   and different types of agents within the group  to be factored in. For present illustrative purposes, we shall simply consider these agent categories to be male and female, but any biological phenotype that influenced natural activity schedules would be appropriate.

\begin{figure}[h!]
	\centering \includegraphics[scale=0.6]{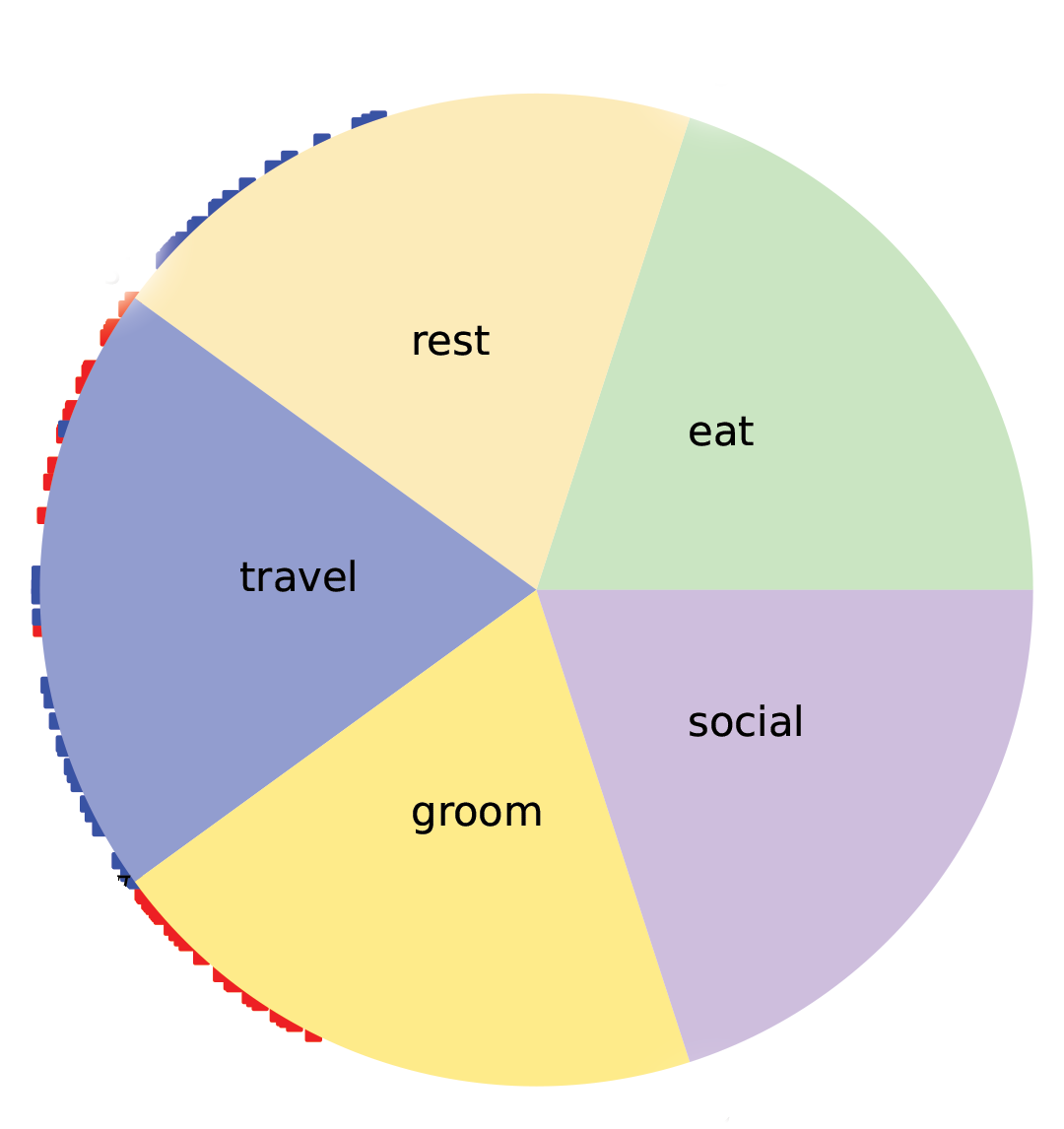}
	\caption{An example of a divided disc with female and make agents doing different activities.}
	\label{example}\label{Figure 1}
\end{figure}

We begin by introducing the background from \cite{[1]} for defining our new model in \autoref{Section II}. Then, in \autoref{Section III}, we modify the agent-based coupled model to study synchrony as animals cycle through various daily activities such as feeding, grooming, and resting. We divide the circle into sectors, and assign a range of angles to each activity, an example of which is shown in \autoref{Figure 1}. In this setting, an agent being in a given range represents it doing the activity defined by that arc. Within our model, we assign each agent an activity to be performing at a certain time, such as eating for 800 units of time and then resting for 700. When an agent moves from one activity to another, we set its desired angle to a randomly selected value in the range. We further divide our group into two categories (Female and Male agents) who move through the activities at different times as a group. After describing the essentials of the model, we study the model under different conditions in \autoref{Section IV}, using two, three, or five sectors as well as the effect of different parameter values on the behavior of the agents. The implications of our model are considered in \autoref{Section V}. First, we use Conradt’s \cite{[5]}  established segregation index to show that:

\begin{itemize}
\item the segregation index is periodic, and
\item the period is similar to the least common multiple of the female and male cycles.
\end{itemize}

Second, we study a coupling value $a_{jl}$ between individuals which takes on values between 0 and 1, for 0 corresponding to the case where agents completely disregard each other. This parameter controls how much agent $j$ cares about information from agent $l$, allowing us to measure female/male segregation by considering the average of $a_{jl}$ where both $j$ and $l$ are female (FF), male (MM), or one male and one female (MF). In particular, we show that by changing the distance agents can detect each other, we cause the MF average to slowly decouple.

\section{Background}\label{Section II}

We here give a short overview of some of the background which will be important for developing our synchrony model through which we study animal coupling allowing for multiple activities to be factored in when different types of agents interact in groups. We build upon the agent-based coupled model developed in \cite{[1]} for one type of agent performing one type of activity.

The agent-based coupled model \cite{[1]} is based on the Kuramoto \cite{[7]} model of coupled oscillators and aims to describe a group of animals making a decision about what direction to head in. The basic premise of this model is that there are three groups of individuals $N_1,~ N_2$ and $N_3$ which have different preferences. Individuals in $N_1$ and $N_2$ are ‘informed’, meaning that they have a preference $(1,2)$ for which direction the group should move in, while individuals in $N_3$ have no preference (‘uninformed’). These models derive from the classic disease propagation models of Kermode $\&$ McKendrick \cite{[G]} where individuals in class $N_3$ are in this case equivalent to the ‘susceptible’ category who are vulnerable to infection by individuals from either of the other two classes.  At every iteration, the angle of every agent is given by the following equations:

 \begin{eqnarray}
 \frac{d\theta_j}{dt} &=& \sin(\bar{\theta_1}-\theta_j(t)) + \nonumber \\
 &~&\frac{K_1}{N}\sum_{l = 1}^N a_{jl}(t) \sin(\theta_l(t) - \theta_j(t)), ~{\rm for }~j \in N_1 , \label{ecu1}\label{eq1}
\\
 \frac{d\theta_j}{dt} &=& \sin(\bar{\theta_2}-\theta_j(t)) + \nonumber \\
 &~&\frac{K_1}{N}\sum_{l = 1}^N a_{jl}(t) \sin(\theta_l(t) - \theta_j(t)),  ~{\rm for }~j \in N_2 , 
\\
 \frac{d\theta_j}{dt} &=& \frac{K_1}{N}\sum_{l = 1}^N a_{jl}(t) \sin(\theta_l(t) - \theta_j(t)),  ~{\rm for }~j \in N_3. \label{ecu2}\label{eq3}
\label{eq:basic}
\end{eqnarray}

 In the previous equations, the parameter $K_1$ gives the amount an agent cares about other individuals’ directions overall versus the direction it wants. In effect, this is the stickiness created by social bonding between individuals that causes an individual to move towards its neighbour’s preference. Moreover, the function $a_{jl}(t)$ describes how much individual $j$ cares about individual $l$ and how much that influences its overall direction. Thus, $K_1$ can be thought of as controlling the overall influence of other individuals while  $a_{jl}(t)$ controls the influence of a single individual. Following \cite{[1]}, the values of the function $a_{jl}(t)$ are then adjusted in every iteration according to the following:
 
 \begin{align}
	\begin{split}
	\frac{da_{jl}}{dt} &= K_2(1 - a_{jl}(t))a_{jl}(t)(\rho_{jl}(t) - r) \\
	\rho_{jl} &= |\cos(\frac{1}{2}(\theta_j - \theta_l))|, 
	\end{split}
\end{align}
 where we introduce the parameter $K_2$ which controls how fast $a_{jl}$ changes, and $r$ represents the distance agents can sense other agents. In other words, the value of $r$ gives the maximum of the distance that the agents can be apart and still be aware of each other: if two agents are more than $r$ apart, they cannot sense each other. In effect, $r$ indexes the stickiness, or gravitational pull, between pairs of individuals -- their likelihood of remaining spatially close enough to each other to be able to determine which activity phase the other individual is in. In primates, this stickiness is created by social grooming \cite{[Dunbar]}.  In these equations, $\rho_{jl}$ represents how aligned the motion of agent $j$ and $l$ are.
 
\begin{figure}[h!]
	\centering \includegraphics[scale=0.6]{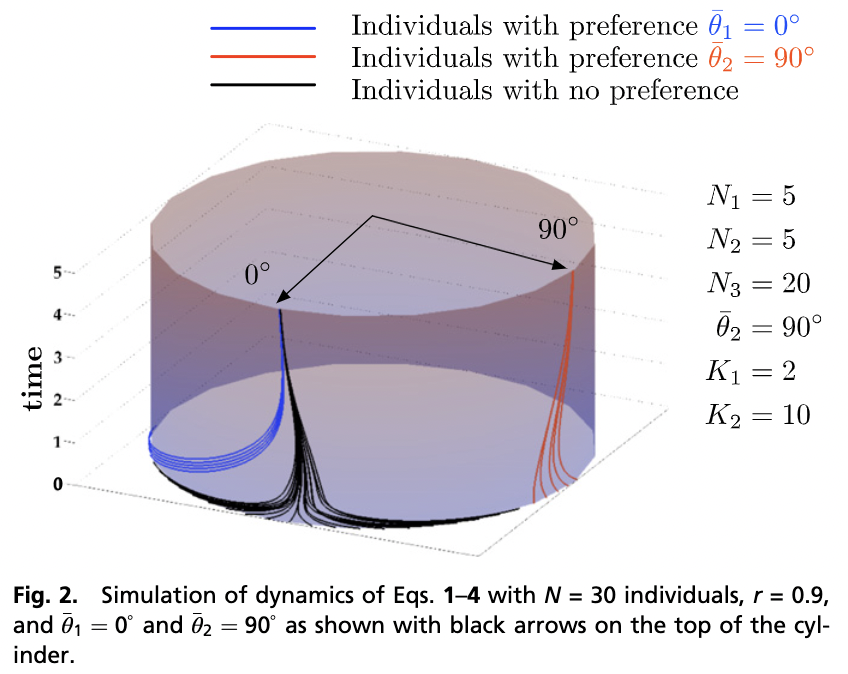}
	\caption{Some of the results of \cite{[1]}.}
	\label{example1}\label{Figure 2}
\end{figure}

 Using a computational agent-based model, it was shown in \cite{[8]} that a group of animals moving together can make a collective decision on direction of travel. The same authors later considered the impact of having informed and uninformed individuals in \cite{[1]} and showed how the dynamics of this system can be modeled analytically. They found that adding uninformed individuals improves stability of collective decision making, where such stability corresponds to most of the group moving in one of two alternative preferred directions. In particular, they showed that this stability depends explicitly on the magnitude of the difference in the preferred directions of each group. An example of their study of the time-continuous system is shown in \autoref{Figure 2}.

In what follows we shall study a setting not encompassed in \cite{[1]}, namely we shall study social groups where two sets of  individuals can have more than two activities available, and where by changing how much individuals care about each other's preferences, the group could decouple.
 
 \section{Feed-rest cycle via a coupled model}\label{Section III}
 
 Expanding on the model \cite{[1]} detailed in the previous section, we consider a system where sectors of the circle represent different activities. For example, an angle between $0$ and $\pi$ could represent feeding, while an angle between $\pi$ and $2\pi$ represents that the animal is currently resting. We shall consider two groups of agents, with all agents being informed individuals with a desired heading. These two categories of agents could be any phenotype or social subnetwork within a group (e.g. a kinship group or a grooming clique), but for convenience (and to relate our analysis directly to the literature on sexual segregation in ungulates) we shall refer to these two groups as males and females. 
 
\begin{figure}[h!]
  \hspace{-0.3 in}
	 \includegraphics[scale=0.3]{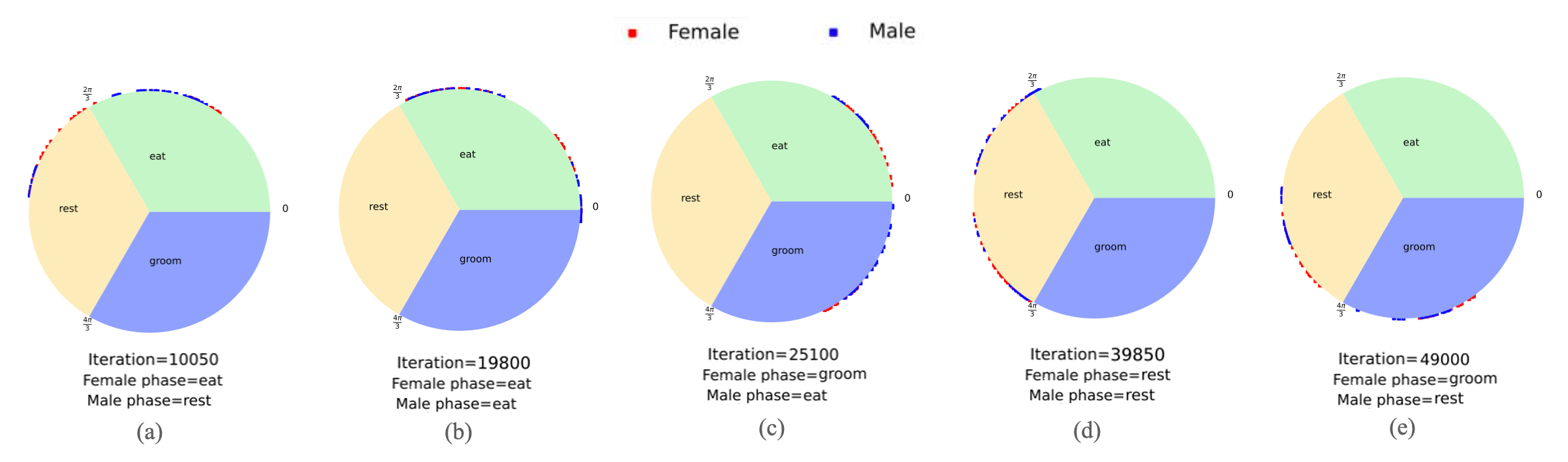}
	\caption{Three sectors for eating, resting, and grooming. 50 males, 50 females. Males spent 400 units of time eating, 600 resting, and 300 grooming. Females spent 300 eating, 700 resting, and 400 grooming. $K_1=2$, $K_2=0.5$.}
	\label{fig:3_7}\label{Figure 4}
\end{figure}

 Based on a schedule, which may be different for males and females, the desired heading of the agent changes to a randomly selected angle within the bounds of the sector. For example, at time 20, males might all desire to move from resting to eating, at which point they would be randomly assigned a desired heading between $\pi$ and $2\pi$. Hence, agents are only randomly assigned a heading when they first move sectors, not every single iteration. In the model, an agent’s current angle can be thought of the activity it is currently performing while its heading is the activity that it would like to be performing. We build the model following Equations \eqref{eq1}-\eqref{eq3} but without any non-informed agents, and adding the separation of the disc as a parameter.
 In what follows we shall describe how our coupled model is defined.  As seen in \autoref{Figure 4}(a), all agents initially start off wanting to rest. 
 
 We randomly set their desired angle at a value between $\pi$ and $2\pi$. After a certain amount of time, m rest time, the males move into the feeding phase, in which case their desired angle is randomly set to a value between 0 and $\pi$. After $f$ rest time, the females also move into the feeding phase. At the end of the day (1000 hours/iterations), all animals' desired angle moves  back to the resting phase. It is important to distinguish
 
 A similar analysis can be performed for agents where three activities are considered, which may represent eating, resting, and grooming, as shown in \autoref{Figure 4}. Females and males spend a fixed amount of time on each activity before moving to a different activity. They spend the same amount of time and perform activities in the same order every iteration. We see that in \autoref{Figure 4}(a), the females and males are typically in different phases/sectors. However, there are still males/females in the ‘wrong’ sector, with more in the incorrect sector compared to the correct one. In \autoref{Figure 4}(b), we see that both groups are in the ‘eat’ sector. Now there seem to be two groups as represented by the two clumps, though they are not necessarily divided by gender. In \autoref{Figure 4}(c), the groups are again mainly in different sectors and once again a lot of them seem to be in a different sector from their desired/scheduled activity. In \autoref{Figure 4}(d), the groups are back together, though still in two clumps but this time the clumps seem closer together. Finally, in \autoref{Figure 4}(e), the groups are supposed to be performing different activities (groom, rest) but a lot of agents are not at the scheduled activity. 
 
\begin{figure}[h!]
   \hspace{0.2 in}
 \includegraphics[scale=0.45]{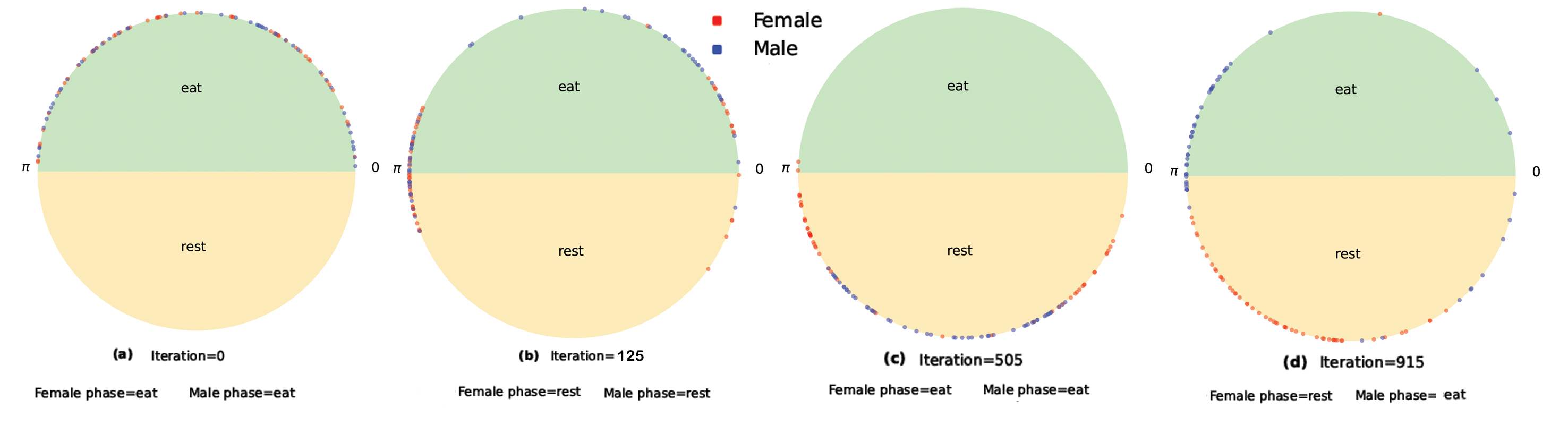}
    \caption{{\bf (a):} initial spawning. {\bf(b):} all agents resting. {\bf (c):} females moving to eating while males stay resting. {\bf (d):} Males foraging while females rest. {\it m rest time} = 400, {\it f rest time} = 450}
	\label{fig:spawn}\label{Figure 3}
\end{figure}

 To understand the impact of considering 5 sectors, we shall study the behaviour of 50 males, 50 females within sectors representing eating, resting, traveling, grooming, and socializing. We shall set males to spend 400 units of time eating, 600 resting, 300 traveling, 200 grooming, and 100 socializing. Equivalently, females shall be set to spend 300 eating, 700 resting, 400 traveling, 300 grooming, and 200 socializing. This schedule is constant and agents continuously cycle between activities in this order and spend the specified amount of iterations in that activity. In order to illustrate the role that the parameter $K_2$ plays, we shall consider different values of K2 from those used in \autoref{Figure 3} and \autoref{Figure 4}. 

\begin{figure}[h!]
\hspace{-0.2 in}
	\centering \includegraphics[scale=0.45]{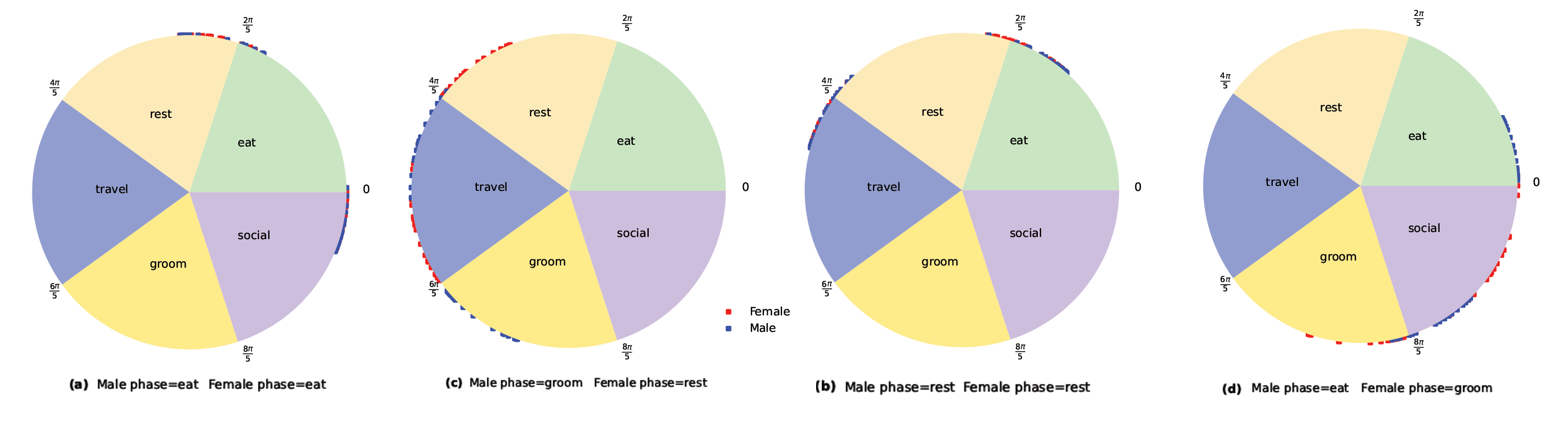}
	\caption{Values of $K_1=2$, $K_2=0.25$ see in different phases and iterations: (a) 50 iterations, (b) 750 iterations, (c) 4550 iterations, and (d) 13100 iterations.}
	\label{fig:5_11}\label{Figure 5}
\end{figure}

With $K_1 = 2$ and $K_2 = 0.25$, we once again see two clumps, except that this time they are spaced so far apart that a significant number of agents are in the wrong sector (\autoref{Figure 5}(a),(b)). However, as the system evolves (\autoref{Figure 5}(c),(d)), we see that most of the females and males are in sectors with a gap between them, with an assortment of agents stuck in the sector between them.

\begin{figure}[h!]
\hspace{-0.2 in}
	\centering \includegraphics[scale=0.45]{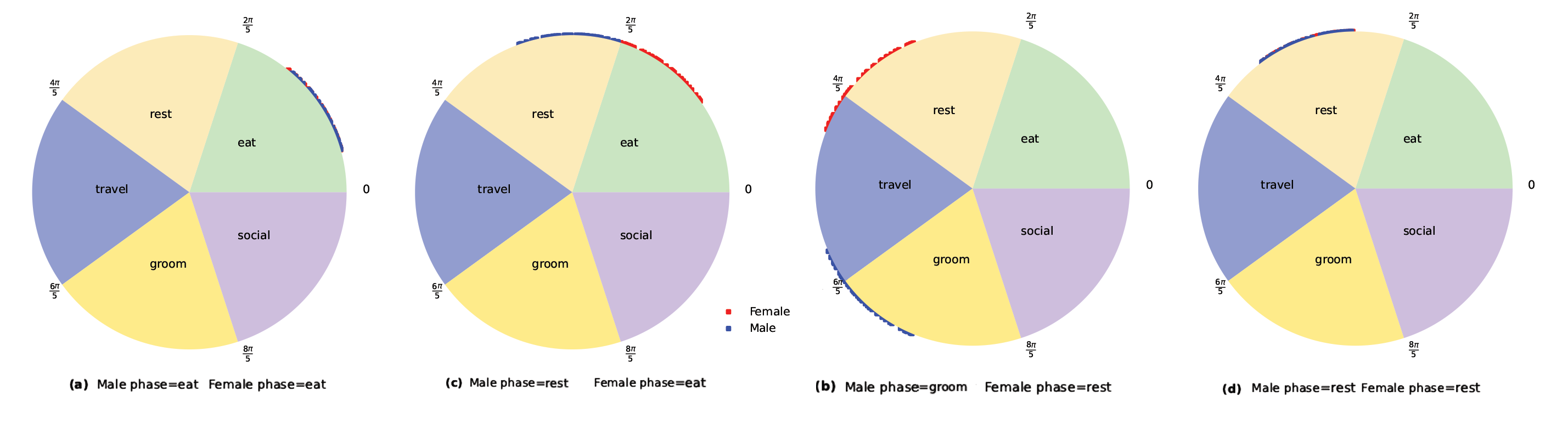}
	\caption{Values of $K_1=2$, $K_2=0$ seen in different phases and iterations: (a) 50 iterations, (b) 4050 iterations, (c) 6300 iterations, and (d) 12100 iterations. }
	\label{fig:5_13_K0}\label{Figure 6}
\end{figure}

With $K_2 = 0$, we find that all the agents initially cluster towards the center of the sector (\autoref{Figure 6}(a)). In \autoref{Figure 6}(b), the males move to resting while the females stay eating and the agents seem to cleanly move towards the right sector with no agents in the wrong sector. In \autoref{Figure 6}(c), the males move to grooming while the females are now resting. There are still some agents in the sector between them (travel) but overall it looks much more clean.

Following the same line as before, in \autoref{Figure 7} below, we consider 50 males and 50 females, with Males assigned to spend 400 units of time eating, 600 resting, 300 traveling, 200 grooming, and 100 socializing; and Females to spend 300 eating, 700 resting, 400 traveling, 300 grooming, and 200 socializing, and $K_2$ set to a high value of 2 (agents care a great deal about the group of agents they are around). It should be noted that the times we have assigned to each activity were inspired by the literature of activity budget studies on a wide range of animals, see for example \cite[Figure 5]{[new]}. In our setting, the  time is set by the successive activity phases that females should be occupying. In \autoref{Figure 7}(a), there are two clumps that are so far apart that most agents are in the wrong sector. In \autoref{Figure 7}(b), the males and females are in different sectors but there appears to be more males than females in the sector the females are actually supposed to be in. In \autoref{Figure 7}(c), the females are supposed to be in rest while the males are in social, but all of them are clumped in the two sectors between rest and social with no agent in the right sector. In \autoref{Figure 7}(d), they are again supposed to be in different sectors but the travel sector where the females are supposed  to be is almost completely filled with males. In \autoref{Figure 7}(e), there are once again no agents in the right sector as they are all in the sectors between them.

\begin{figure}[h!]
 \hspace{-0.2 in}
	\centering \includegraphics[scale=0.55]{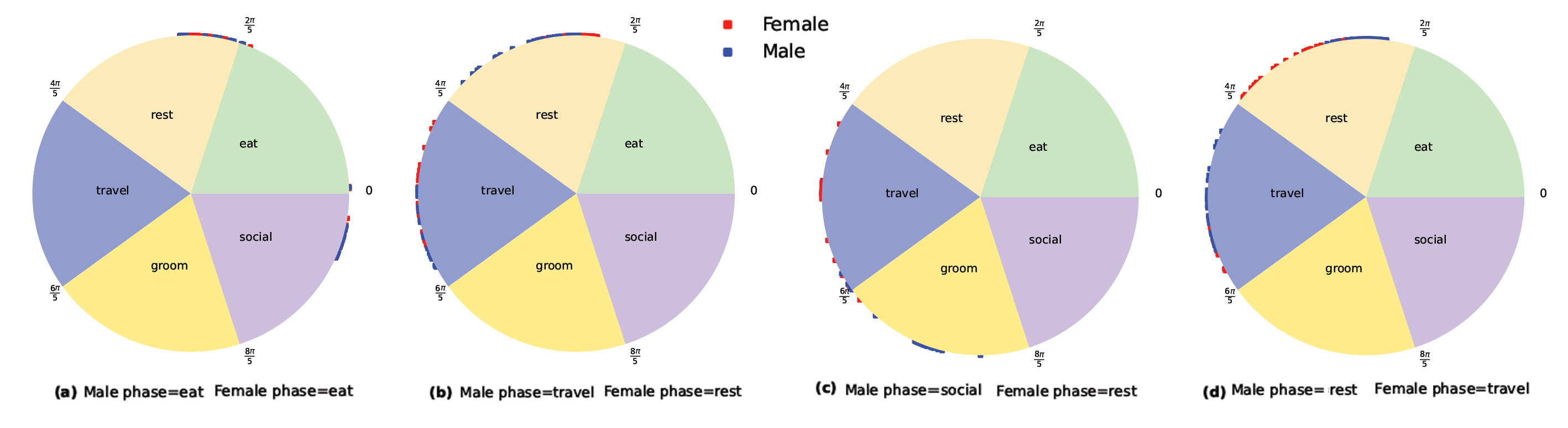}
	\caption{ Values of $K_1=2$ and $K_2=2$ seen in different phases and iterations: (a) 1950 iterations, (b) 2850 iterations, (c) 6400 iterations, and (d) 7100 iterations.}
	\label{fig:5_15}\label{Figure 7}
\end{figure}

\section{Subgroups}\label{Section IV}
\begin{figure}[h!]
 \hspace{-0.2 in}
	\centering \includegraphics[scale=0.4]{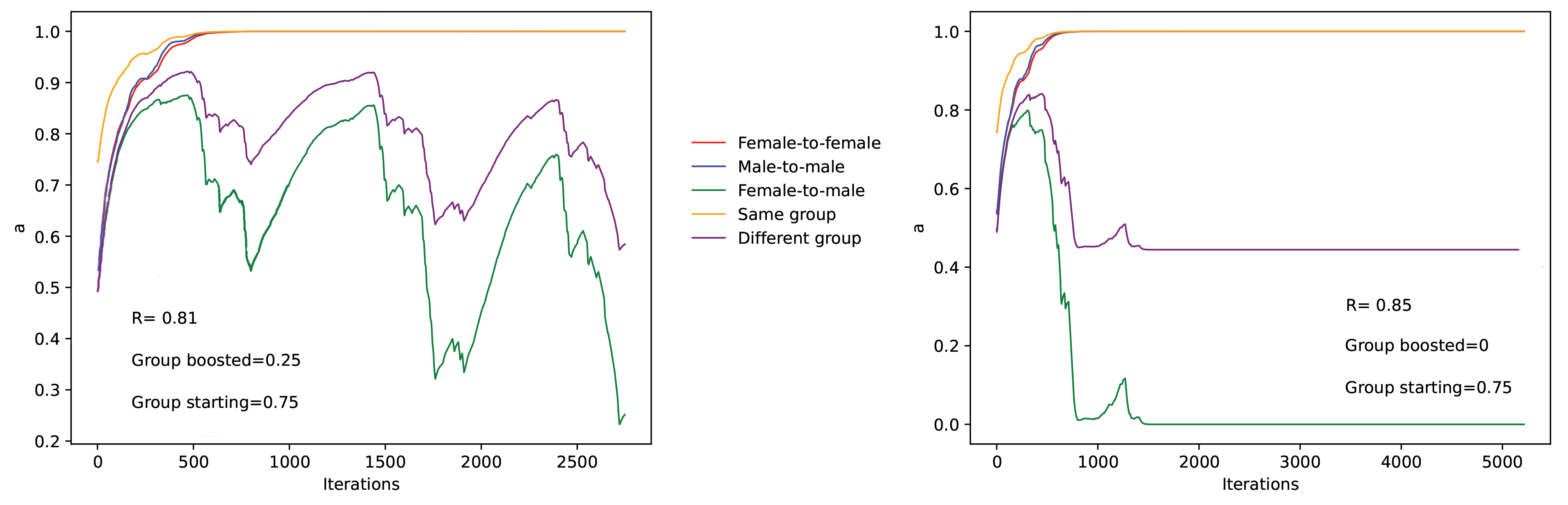}
 	\caption{Left: graph of all $a$ averages with $r = 0.81$, group boosted $= 0.25$, group starting $=0.75$. Right: $r = 0.85$, boosted = 0, group starting = 0.75.}
	\label{fig:33}\label{Figure 8}
\end{figure}

In order to understand how female and male groups interact throughout the course of time, in what follows we create multiple groups of females and males (5 groups with 10 females each and another 5 groups of males). To accommodate our groups, we tweak
the model in two ways:
\begin{itemize}
\item the initial a value between members of the same group ($g_{starting}$), and
\item  adding $g_{boosted}$ to a leading to the following:
\begin{eqnarray}
	&\frac{d\theta_j}{dt} &= \sin(\bar{\theta_1}-\theta_j(t)) \label{eq:groupboosted} \label{(5)}\\&+& \frac{K_1}{N}\sum_{ j \in N_1,l = 1}^N (a_{jl}(t) + group_{boosted})\sin(\theta_l(t) - \theta_j(t)) \nonumber
\end{eqnarray}
\end{itemize}
From a visual perspective, we use differently shaded dots to represent different groups. Just like the MF, FF, and MM a average, we take the average of $a_{jl}$ for where $j$ and $l$ are in the same group and also in different groups. In \autoref{Figure 8}, we see that the same group average goes very quickly to 1 which is to be expected given that we are boosting it and starting it higher. Moreover, one can also see that the average between different groups seems to follow the FM average.

In order to measure how close together a given group is, we use a spatial version of standard deviation where we first convert the agents’ positions into rectangular coordinates, compute the midpoint, and then use the distance from each point to the midpoint.  

\begin{figure}[h!]
 \hspace{-0.2 in}
	\centering \includegraphics[scale=0.4]{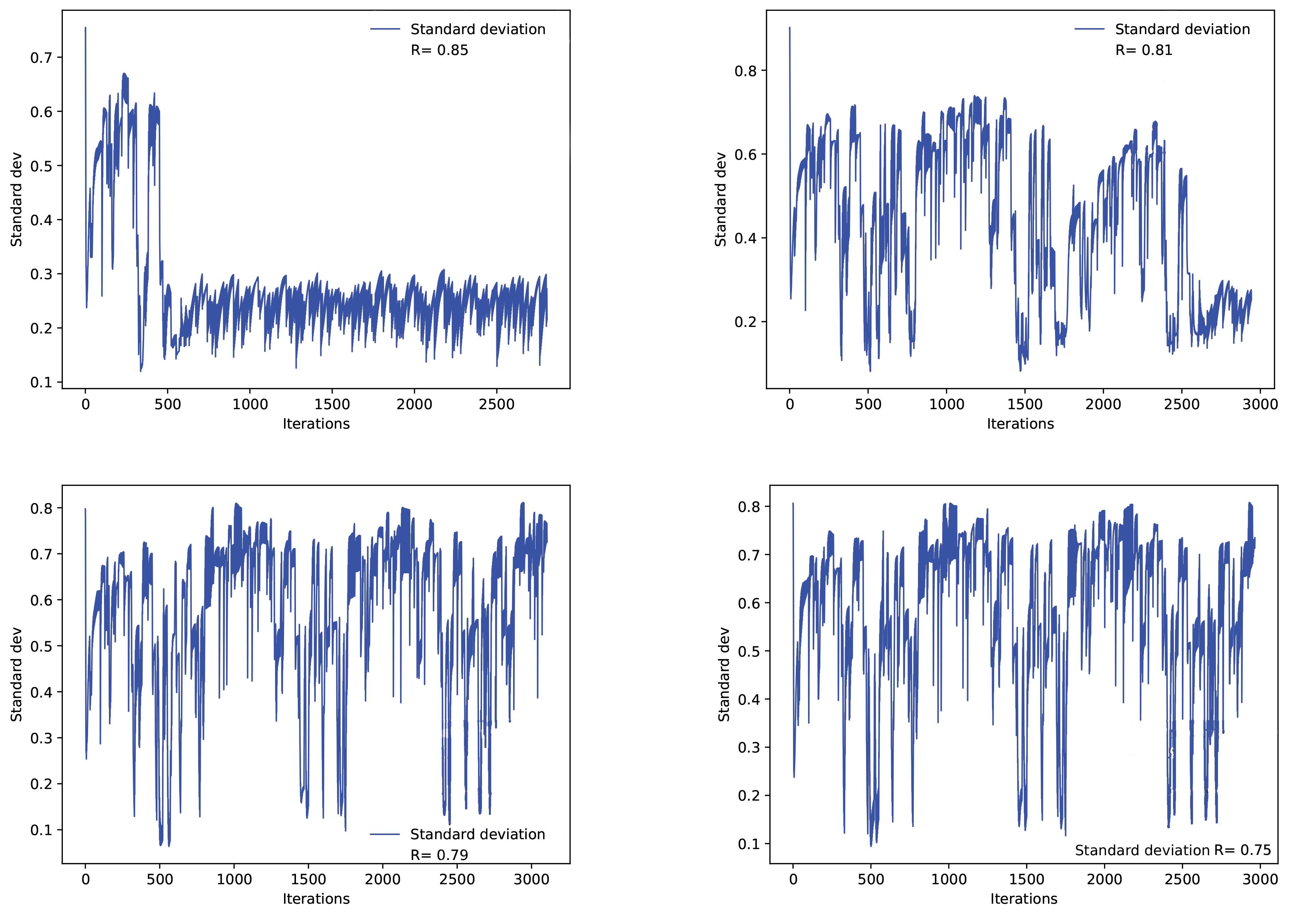}	
	\caption{Graph of standard deviation with varying $r$ values. From top left to bottom right, $r=0.85, 0.81, 0.79, 0.75$. boosted = 0.5, starting = 0.75 for all trials.}
	\label{fig:stdev}\label{Figure 9}
\end{figure}

We first examine the effect of the $r$ value on the standard deviation, and see that a higher $r$ value of 0.85 and 0.81 results in a lower standard deviation although there is less of a visible change between $r = 0.79$ and $r = 0.75$, as shown in \autoref{Figure 9}.

Within the model we introduced above, the desired angle $\theta$ is assigned to a random number in the sector range; as a result, there is a large range of desired angles and this was keeping agents from the same group apart. Hence, in order to improve our model to encode subgroups, we modify Equation \eqref{(5)} to introduce another parameter, $K_0$ allowing us to decrease the importance of the desired angle. In this modified model, $K_0$ must strike a balance between still keeping agents going to the correct sectors/activities but also allowing them to group:
\begin{eqnarray}
	\frac{d\theta_j}{dt} &=& K_0\sin(\bar{\theta_1}-\theta_j(t)) \label{(6)}\\
	&+& \frac{K_1}{N}\sum_{l = 1}^N (a_{jl}(t) + group_{boosted})\sin(\theta_l(t) - \theta_j(t)).\nonumber
\end{eqnarray}

\begin{figure}[h!]
 \hspace{-0.2 in}
	\centering \includegraphics[scale=0.35]{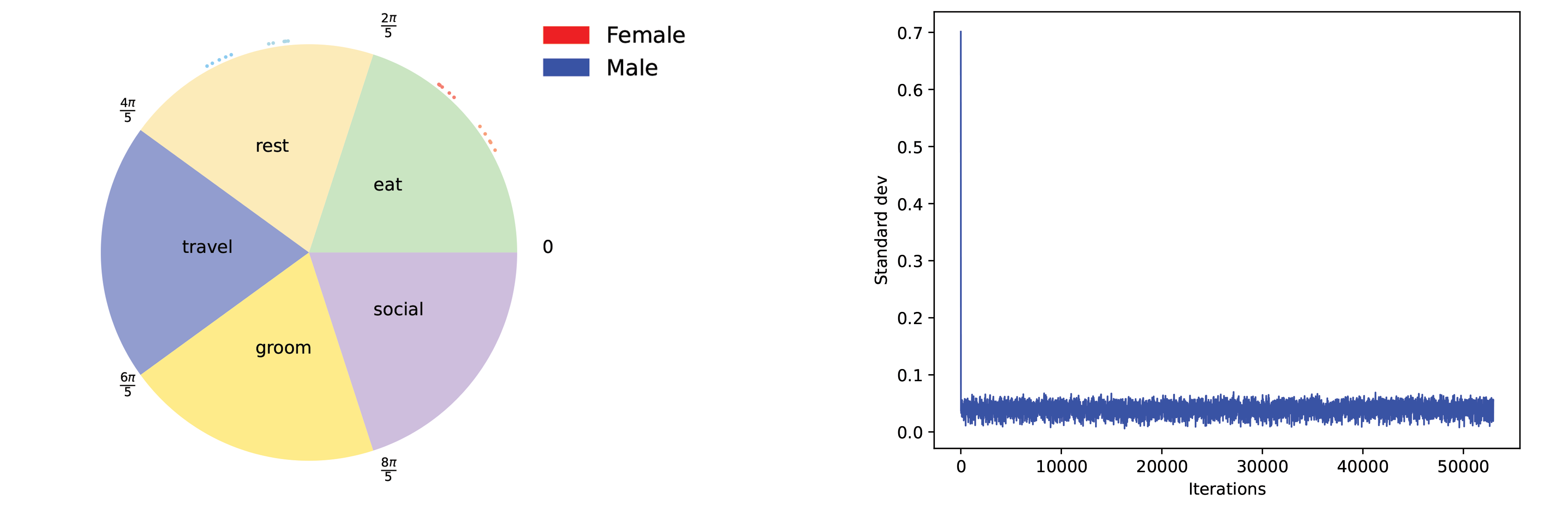}
	\caption{Study of model for $K_0,  K_1, K_2=0.1$.  Female phase=eat and Male phase=rest; R=0.85, Boosted$=40$, starting$=1$. Left: Iteration=2485, MF average=0.00011244701793142417, Std dev average=0.04071990250949034.}
	\label{fig:89_2485}\label{Figure 10}
\end{figure}

In order to understand the relevance of $K_0$, we begin by considering 2 groups of females and 2 groups of males with only 5 agents per group. Through our model, we can see the agents stick to their own groups, as depicted in \autoref{Figure 10}. An additional key observation we make is the balance between $K_1$ and boosted. By decreasing the values of $K_1$ and proportionally increasing boosted, we can decrease the influence of agents outside the group while keeping the influence of agents within the group the same. We also see in the line chart that the standard deviation of the agents is lower than in \autoref{Figure 9}.

\begin{figure}[h!]
 \hspace{-0.2 in}
	\centering \includegraphics[scale=0.35]{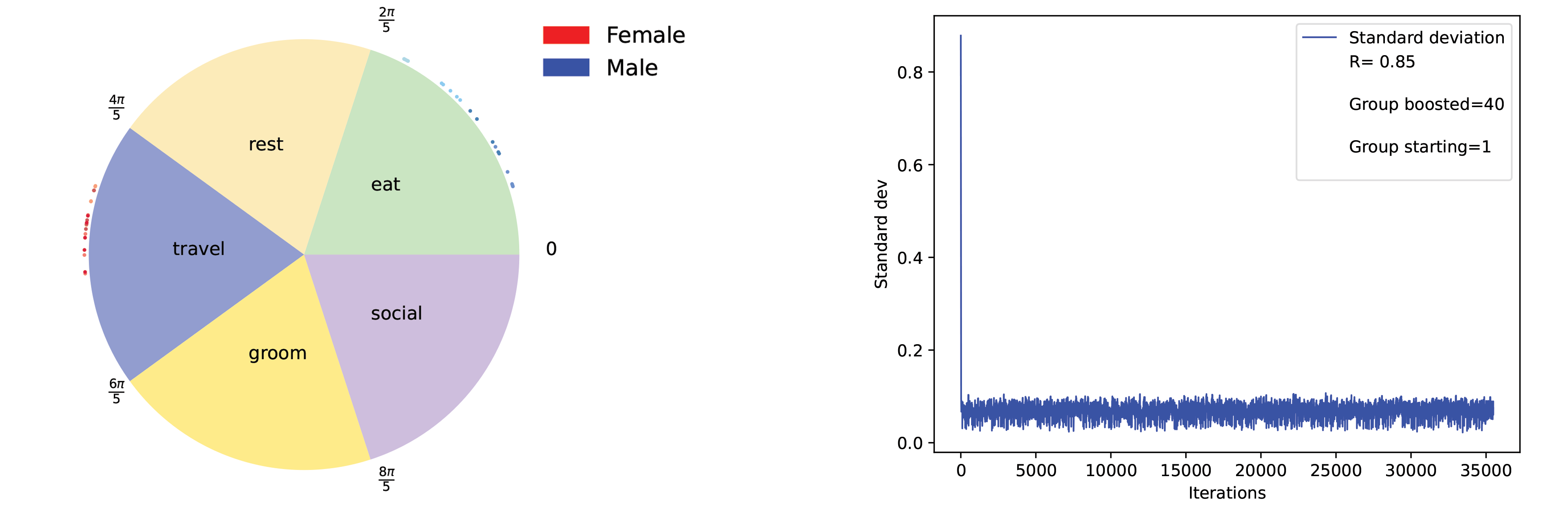}
	\caption{Study of model for $K_0,  K_1, K_2=0.1$.  Female phase=travel and Male phase=eat; R=0.85, Boosted$=40$, starting$=1$. Left: Iteration=1635, MF average=0.0001130389055524958, Std dev average=0.06870473883390149.}
	\label{fig:91}\label{Figure 11}
\end{figure}

The effect of making both the number of subgroups and their sizes larger is shown in \autoref{Figure 11}, where we consider 4 groups of each gender and 40 total agents. Notice that the male agents form distinct groups but the female groups seem to overlap rather more with each other. This suggests that groups being visually close might be dependent on their randomly chosen desired angle. Finally, comparing the results shown in \autoref{Figure 10} and \autoref{Figure 11}, it should be noted that by considering the standard deviation values one can see that group members are overall much closer to each other, as shown in \autoref{Figure 10}.

\begin{figure}[h!]
 \hspace{-0.2 in}
	\centering \includegraphics[scale=0.35]{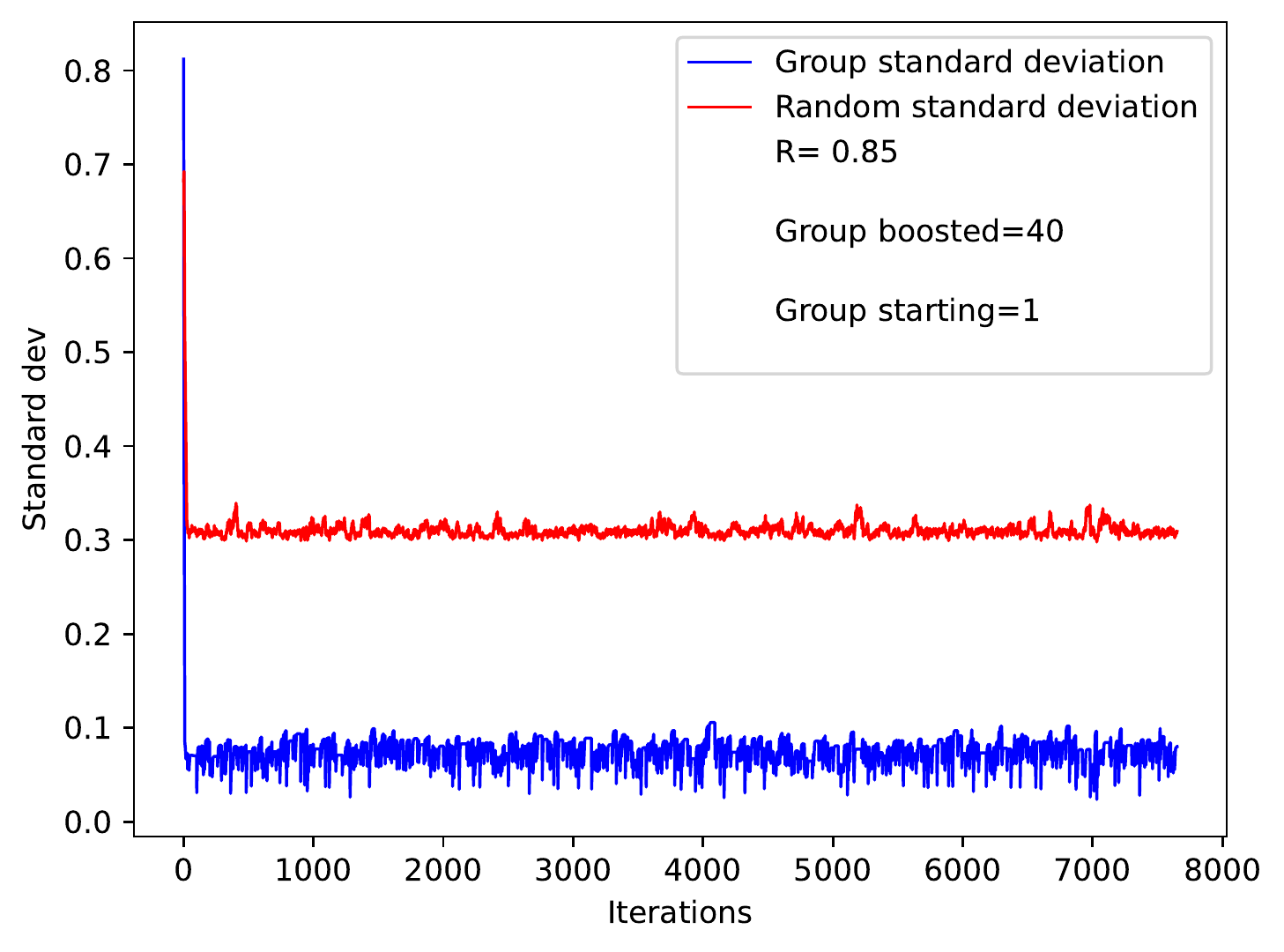}
	\caption{Standard deviations for $K_0=0.1, K_1=0.1, K_2=0.1. $ boosted$=40$, starting$=1$.}
	\label{fig:5_both}\label{Figure 12}
\end{figure}

We conclude this section by comparing the standard deviation between members of a group and randomly selected members of the same gender. For the randomly selected members, each sample shall consist of the same number of individuals as the group size. We perform our study by considering the following: we looked at two times the number of group samples taken, with half of them being random samples of the male agents and the other half being random samples of the female agents. As we can see from \autoref{Figure 12}, there is a large difference between the standard deviation of members  in the same group and those who were not. Thus, one can conclude that our grouping model is consistent and functional. 
In what follows, we briefly discuss the implications both of our original model as well as the extension model which allows for subgroups of agents of the same type.

\section{Discussion}\label{Section V}The spatial segregation index is a commonly used metric in animal behaviour studies \cite{[5]}. In \autoref{Section VA} below, we analyze the spatial segregation within our model by computing the index in every iteration using the number of males and females in each given sector of the circle. Graphing the spatial segregation index, we find that it appears to be periodic and have a similar period to the least common multiple of the female and male cycles.

The coupling parameter $a_{jl}$ between individuals is then introduced in the framework of our model, which controls how much agent $j$ cares about information from agent $l$. This value measures female/male segregation. For this, we  took the average of $a_{jl}$ where both same sex and mixed sex groupings, referring to these as the MM, MF, and FF average values. In particular, we show that changing $r$, the distance agents can detect each other, can cause the MF average to slowly decouple.

\subsection{Synchrony index}\label{Section VA}
\newcommand{\scsp}{SC_{\text{spatial}}}

\begin{figure}[h!]
 \hspace{-0.2 in}
	\centering \includegraphics[scale=0.2]{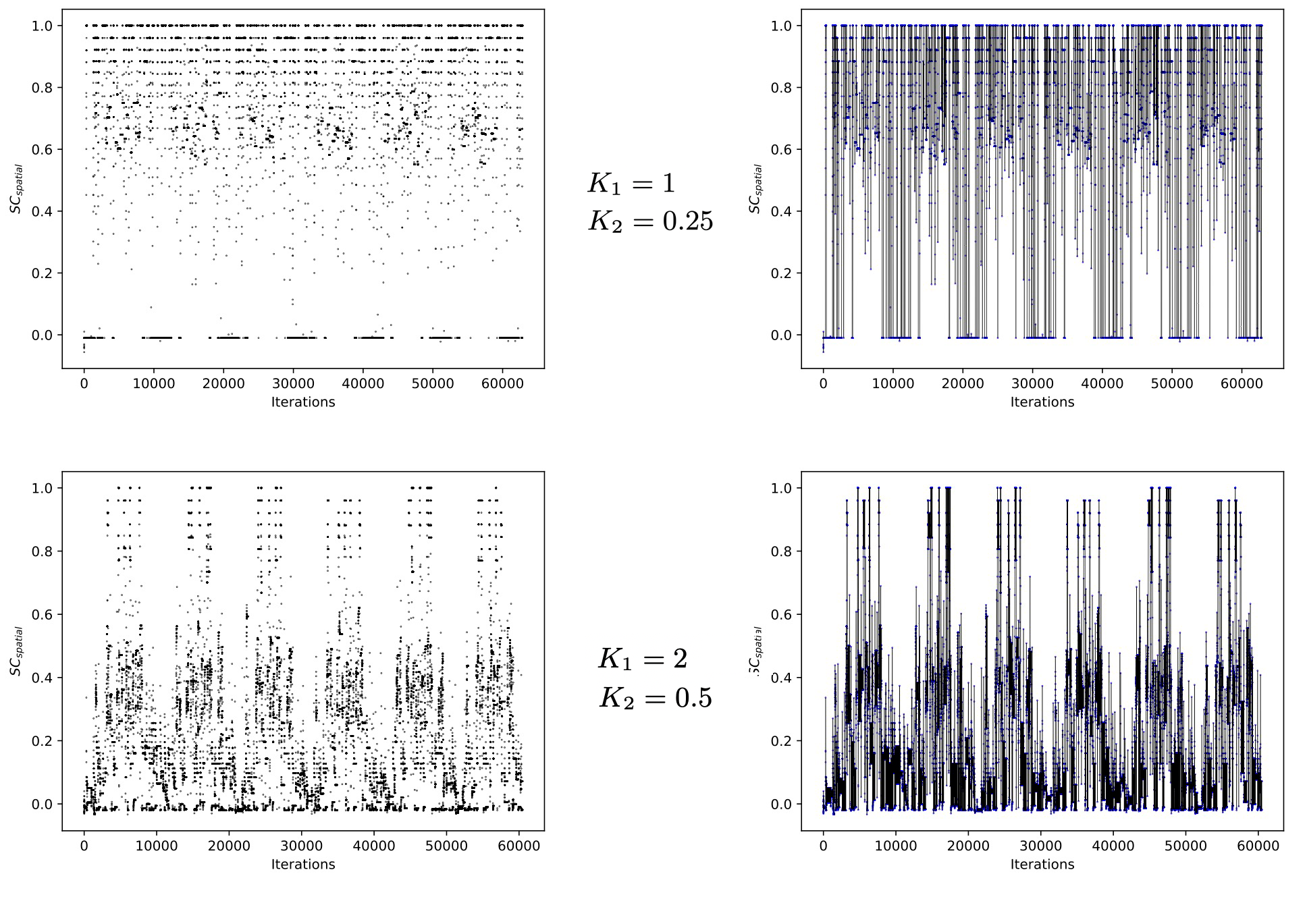}
	\caption{Plots done for 50 males, 50 females. Males spent 400 units of time eating, 600 resting, 300 traveling, 200 grooming, and 100 socializing. Females spent 300 eating, 700 resting, 400 traveling, 300 grooming, and 200 socializing. Top:  $K_1=1$, $K_2=0.25$. Bottom: $K_1=2$, $K_2=0.5$}
	\label{fig:1-0.25-scatter}\label{Figure 13}
\end{figure}

In order to study the properties of our model further, we shall consider the spatial segregation index
\begin{equation}
\scsp:= 1 - \frac{C}{A\cdot B} \cdot \sum_{i = 1} ^ r \frac{a_i \cdot b_i}{c_i - 1} ,
\label{eq:sc_spatial}
\end{equation}
where $a_i$ is the number of males in the $i$-th circle sector, $b_i$ is number of females in $i$-th sector, $c_i = a_i+b_i$, and  $A$ is number of males, $B$ is number of females, the value of $C = A+B$ and we exclude animals in their own sector. The values of $SC_{spatial}$ can be between 0 to 1, where 0 represents no segregation and 1 represents complete segregation. By calculating $SC_{spatial}$ over many iterations and graphing it as a scatterplot (\autoref{Figure 13}), we can see from the upper pair of images that it always seems to hit certain $SC_{spatial}$ values as the points appear to be in rooms. In the lower pair of images, we can see a line plot where $SC_{spatial}$ appears to be periodic, possibly due to the different length of female and male cycles, or how their cycles are offset.

The distribution of points in the upper pair of panels of \autoref{Figure 13} suggests that the  points are more likely to appear in rows that are either close to a $SC_{spatial}$ value of 1.0 or close to 0.0, with the former being more common. \autoref{Figure 13} also suggests that there is a tendency for a periodicity in the distribution of the data points. In the lower pair of panels, $SC_{spatial}$ appears to suddenly spike up to 1.0 before falling again, creating a distinctly W-shaped line. This is likely to be due to males and females moving into and out of phase, causing $SC_{spatial}$ to change abruptly. In this version of the model, males have a 1600 unit cycle while females have a 1900 unit cycle. The Least Common Multiple of 1600 and 1900 is 30400 which closely matches the periodicity of the graphs in \autoref{Figure 13}.

\begin{figure}[h!]
 \hspace{-0.2 in}
	\centering \includegraphics[scale=0.55]{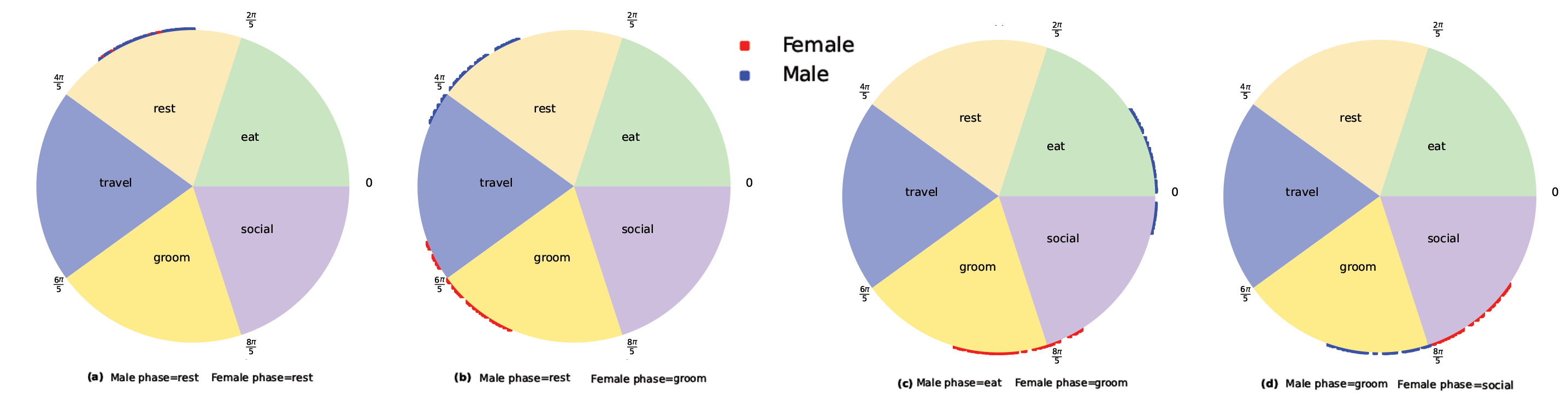}
	\caption{Values of  $K_1=1$, $K_2=0.5$ seen at different iterations: (a) 500 iterations  with $
SC_{spatial}=-0.010101010101010166$, (b) 7350 iterations with $SC_{spatial}=0.7495652173913043$, (c) 13100 iterations with $SC_{spatial}=0.7312000000000001$, and (d) 18950 iterations with $SC_{spatial}=0.9215686274509804$.}
	\label{fig:30_combined}\label{Figure 14}
\end{figure}

\begin{figure}[h!]
 \hspace{-0.2 in}
	\centering \includegraphics[scale=0.55]{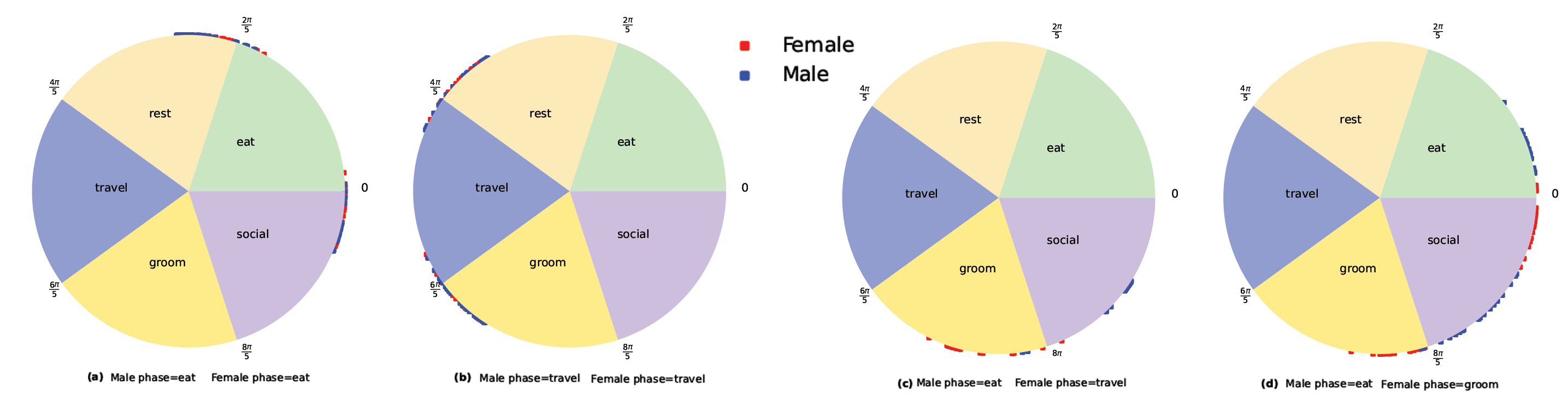}
	\caption{ Values of $K_1=2$, $K_2=0.5$ seen at different iterations:   (a) 150 iterations  with $
SC_{spatial}=-0.02578754578754583$, (b) 1250 iterations with $SC_{spatial}=0.045459459459459395$, (c) 3250 iterations with $SC_{spatial}=0.8815999999999999$, and (d) 3500 iterations with $SC_{spatial}=0.46090909090909093$. }
	\label{fig:36_combined}\label{Figure 15}
\end{figure}

In \autoref{Figure 14} and \autoref{Figure 15}, we considered 50 males and 50 females as before. Males are set to spend 400 units of time eating, 600 resting, 300 traveling, 200 grooming, and 100 socializing. Females set to spend 300 eating, 700 resting, 400 traveling, 300 grooming, and 200 socializing. In \autoref{Figure 14} (a), we see that the females and males do a better job of going to their assigned sector due to the lower K1 value, and observe much higher $SC_{spatial}$ values than in \autoref{Figure 14} (b), (c) and (d). Then, in \autoref{Figure 15} we see the progression of a trial. Initially in \autoref{Figure 15} (a), the groups are very mixed and have a low $SC_{spatial}$, but as they move into different phases in \autoref{Figure 15} (c) and (d), the value of $SC_{spatial}$ increases.

\newpage
\subsection{Measuring the coupling value}

Recall that the coupling value $a_{jl}$ measures of how much weight individual $j$ places on information from individual $l$, where $a_{jl}$ takes on values between 0 and 1, for 0 corresponding to the case where agents completely disregard each other. 
Using the definition of $a_{jl}$ in Equation \eqref{(88)}, one can deduce how to adjust the parameter $r$ as follows:
\begin{equation}
	\frac{da_{jl}}{dt} = K_2(1 - a_{jl}(t))a_{jl}(t)(\rho_{jl}(t) - r).\label{(88)}
\end{equation}
We test various values for $r$ and examine its impact on $a_{jl}$.  To do this, we compute the average value of a between male and male agents, female and female agents, and male and female agents.

\begin{figure}[h!]
 \hspace{-0.2 in}
	\centering \includegraphics[scale=0.3]{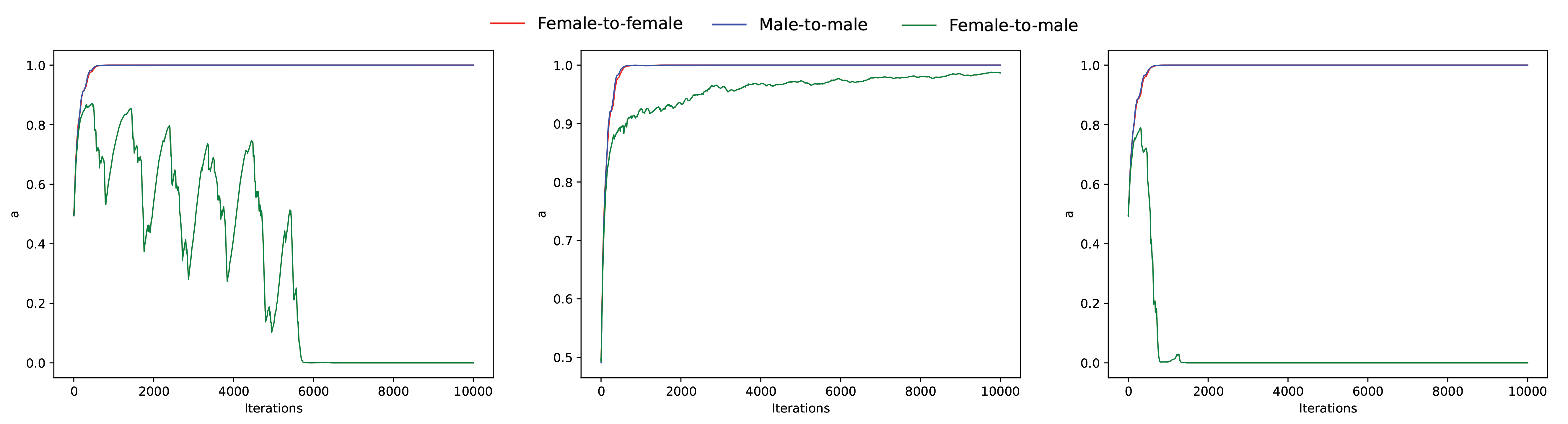}
			\caption{A long, cyclic decoupling process. As before, we have 50 males and 50 females. Males spend 40 units of time eating, 60 resting, 30 traveling, 20 grooming, and 10 socializing. Females spend 30 units of time eating, 70 resting, 40 traveling, 30 grooming, and 20 socializing. Note that this is scaled down by 10 from previous experiments to save computational time. Values of  $K_1 =  2, K_2 = 0.1$ for   (Left) $r=0.817$, (Middle) $r=0.81$ and (Right)  $r=0.85$.} 
			\label{fig:817_a}\label{Figure 16}
\end{figure}

 Looking at the decoupling process, one can see that $r = 0.817 $ leads to the longest decoupling process, and thus we consider this setting in \autoref{Figure 16} (left). We start all as at 0.5, and see the male-male and female-female a average increase to 1 quickly, as is expected. Moreover, we see the female-male average initially grow to above 0.8 before decreasing and increasing in cycles as their phases go in and out of synchrony. At approximately 6000 iterations, we see the male and female groups completely decouple. In \autoref{Figure 16} (middle), we use $r = 0.81$ and see a gradual increase in the average coupling value for females-to-males. The coupling value appears to increase rapidly in the beginning as $(1-a_{jl})a_{jl}$ is maximized with $a_{jl} = 0.5$ and in the beginning the phases of males and females are relatively in synchrony. We then see periodic dips and increases similar to that in \autoref{Figure 16} (left) except with much smaller amplitude and period. Overall, there is an upwards trend and appears to begin to approach 1.0. Based on additional trials with $r$, we can conclude that values of $r$ less than or equal to 0.81 will not result in decoupling but instead a convergence to 1.

In \autoref{Figure 16} (right), we use a value of $r=0.85$. We see the female-to-male average increase to slightly below 0.8, which is lower than in \autoref{Figure 16} (left). We then see it quickly decouple to an average 0. Although there are some dips, there does not appear to be a clear cycle. In addition, it is worth noting that the female-to-female as well as male-to-male averages take longer to converge to 1. Using information from other trials, we conclude that the greater $r$ gets, the faster the female-to-male average decouples to 0. 

\begin{figure}[h!]
 \hspace{-0.2 in}
	\centering \includegraphics[scale=0.6]{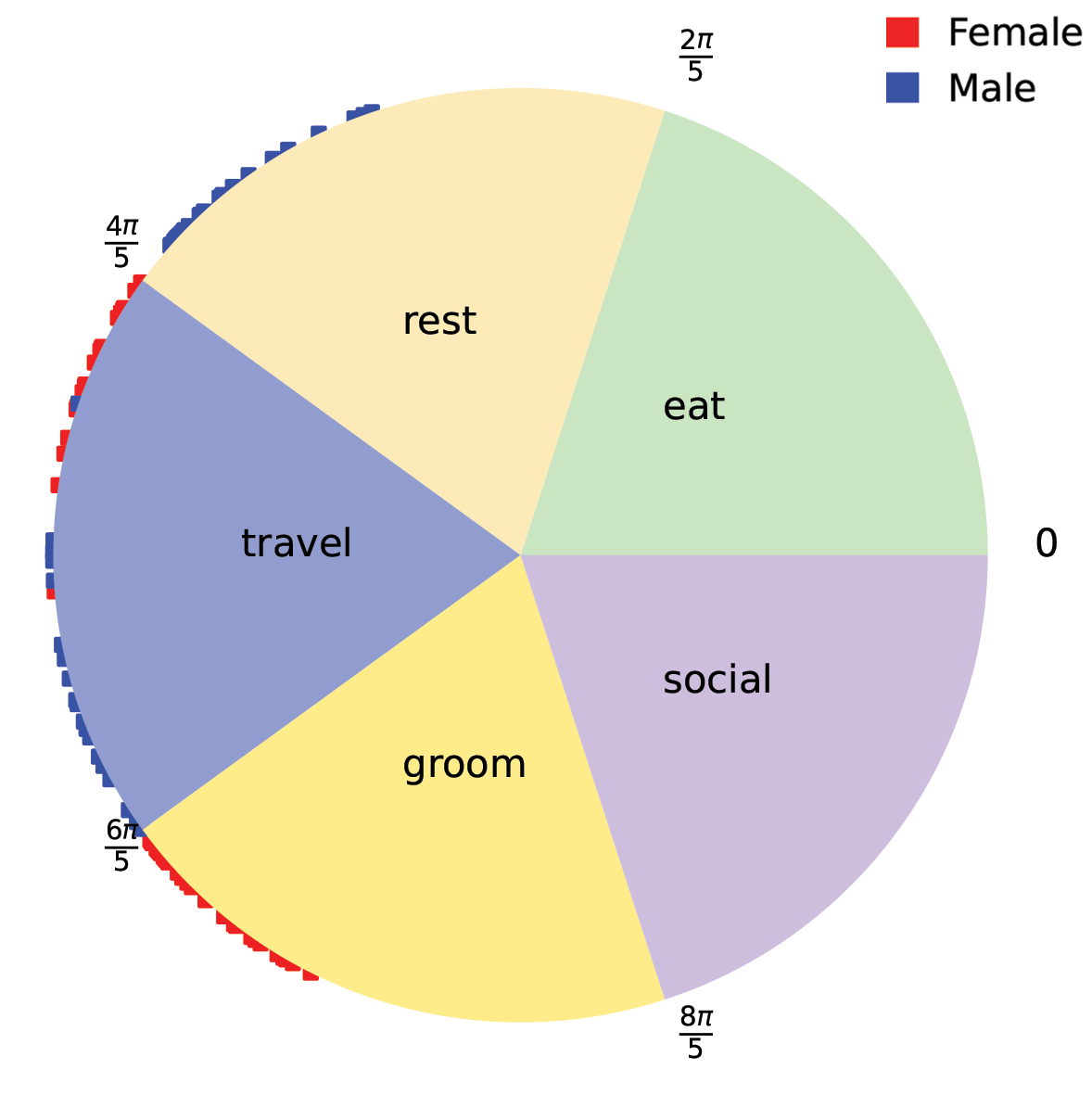}
	\caption{Parameters set as in   \autoref{fig:817_a} (middle). MF average=0.9631973315319021, Iteration=3585
Female phase=groom, Male phase=rest}
	\label{fig:coupled_5}\label{Figure 17}
\end{figure}

\begin{figure}[h!]
 \hspace{-0.2 in}
	\centering \includegraphics[scale=0.6]{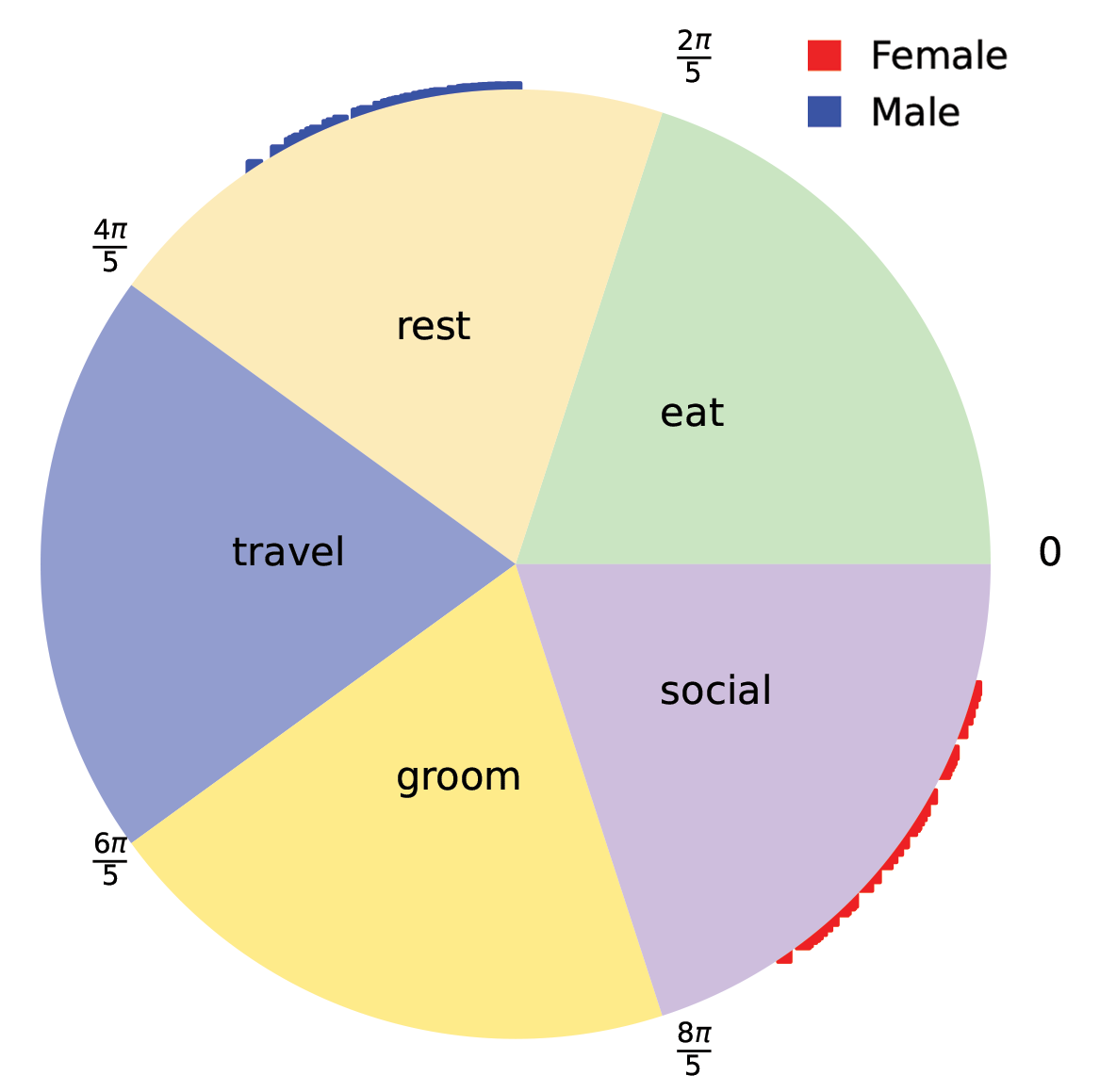}
	\caption{All parameters besides $r$ are the same as in Figure \autoref{fig:817_a} (right). MF average=1.1132096604960067e-08, Iteration=1700
Female phase=social, Male phase=travel}
	\label{fig:decoupled_5}\label{Figure 18}
\end{figure}

In \autoref{Figure 17}, we see that, within the setting of \autoref{Figure 16} (middle), the agents have a MF average close to 1 and are therefore very closely coupled. As a result, they do not go to the correct phases but instead there is large overlap in the phases between states such as travel. Within the setting of \autoref{Figure 16} (right), in \autoref{Figure 18}, we see that there is a low male-female average close to zero. As a result, the male and female groups are greatly decoupled. We can see that they go to their assigned sector or activity with little regard to each other.

\begin{figure}[h!]
 \hspace{-0.2 in}
	\centering \includegraphics[scale=0.5]{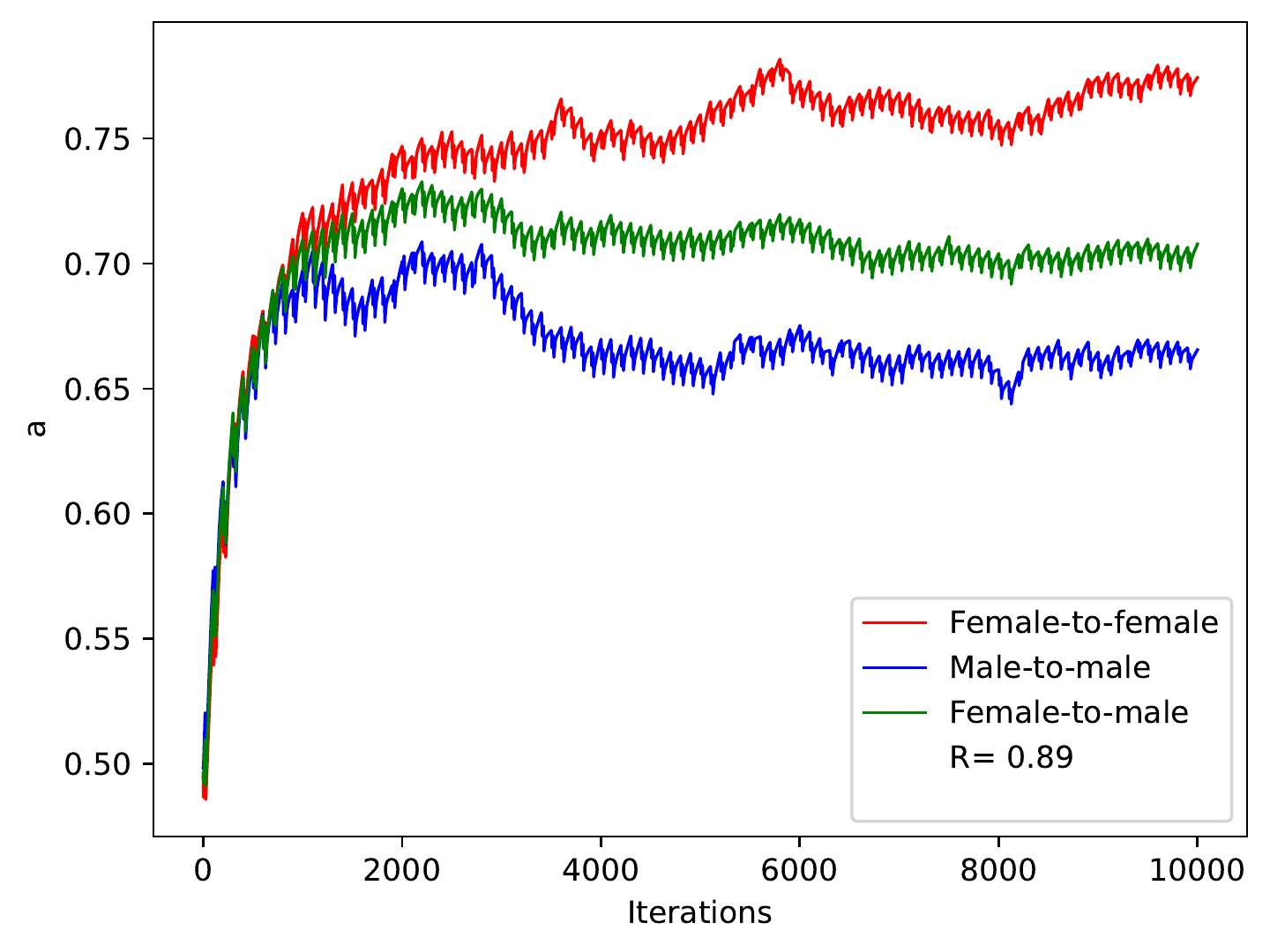}
	\caption{  Males spend 19 units of time eating and then 81 resting while females spend 21 eating and then 79 resting. Parameters set to $K_1 = 2, K_2=0.1$ and $r=0.89$.}
	\label{fig:two_phase_bad}\label{Figure 19}
\end{figure}

It is interesting to understand our model within a two-state setting, with just rest and forage and use a ratio of approximately 20 move and 80 stay. For this, we consider a model where males spend 19 units eating and 81 units rest, while females spend 21 units eating and then 79 units resting. This setting results in the same cycle length for both males and females (\autoref{Figure 19}).

Notice that the female-to-male synchrony is actually higher than that of the male-to-male. For everything except for two iterations, the males and females are all in the same sector, so one would expect FF, MM, and MF to all have the same synchrony. It seems that when the males first move from eating to resting while the females remain resting, they end up getting scattered.

Since the sectors are so large, two agents that are in different sectors could be much more in synchrony than those who are in the same sector. To help account for this,
we add in a multiplier $m$ in Equation \eqref{(99)}, in such a way that if the agents are not in the same sector, $m$ is lower, leading to the following:
\begin{equation}
	\frac{da_{jl}}{dt} = K_2(1 - a_{jl}(t))a_{jl}(t)(m\rho_{jl}(t) - r). 
	\label{(99)}
\end{equation}

\begin{figure}[h!]
 \hspace{-0.2 in}
	\centering \includegraphics[scale=0.55]{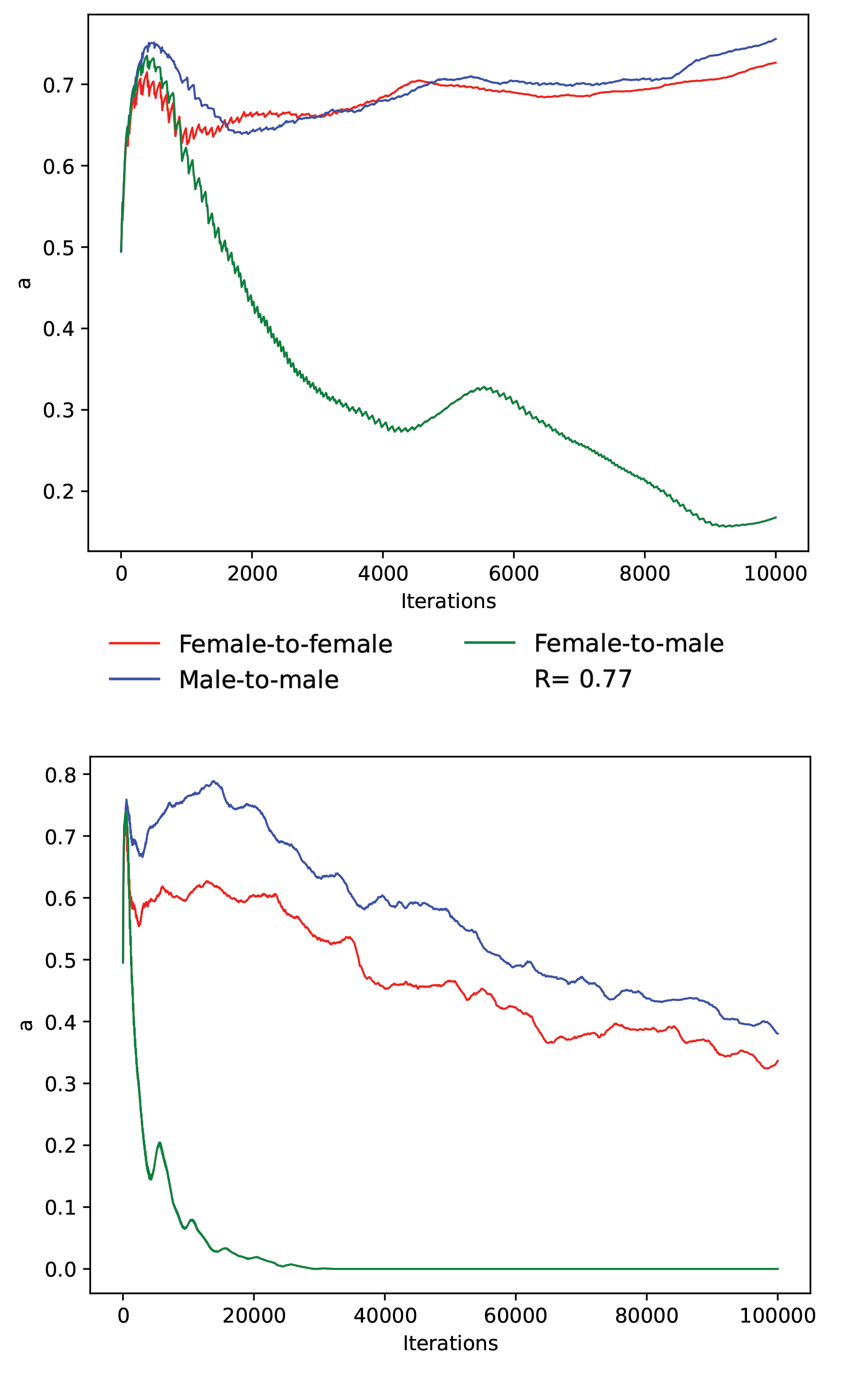}
	\caption{ (Top) $r=0.77$. Males spend 21 units of time eating and then 80 resting while females spend 19 eating and then 80 resting. $K_1 = 2, K_2=0.1$. $m_{\text{same sector}} = 1, m_{\text{different sector}} = 0.8.$; (Bottom) $r=0.77$. Males spend 21 units of time eating and then 80 resting while females spend 19 eating and then 80 resting. $K_1 = 2, K_2=0.1$. $m_{\text{same sector}} = 1, m_{\text{different sector}} = 0.8.$}
	\label{fig:2_16}\label{Figure 20}
\end{figure}

In \autoref{Figure 20} (top), we successfully have the female and male groups decoupling while male-male and female-female stay coupled. It is worth noting that the male and female lines appear to hover around 0.7. It seems unlikely that they will ever completely couple to 1.0 considering that each sector is so large. In \autoref{Figure 20} (bottom), we use the same parameters (but not the same seed) as in \autoref{Figure 20} (top) but run it for 10 times as long, and see that, after initially peaking, MM and FF synchrony decreases.

\section{Conclussions}
In natural primate social groups, animals risk the fragmentation of groups during daily foraging if individuals' activity schedules become desynchronised. To avoid this, species need strategies that increase the stickiness, or gravitational pull, that causes individuals to coordinate their activities. Failure to do so results in groups fragmenting and hence the loss of the benefits gained from living in large social groups. Our analyses indicate that the more disparate the activity schedules of group members 
(here simplified to two phenotypes, males and females), the more likely groups are to fragment. Furthermore, this tendency is exacerbated by lower values of $r$, the stickiness between individuals. Our concern here has been to establish proof of principle rather than to establish exactly what magnitude of these values causes a phase transition between a cohesive group and a fragmented one. Nonetheless, our analyses are sufficient to show that the transition between a single coherent group and a fragmented one occurs quite suddenly in a distinction phase transition. Since differences in activity schedules are usually determined by differences in body mass or reproductive state (e.g. pregnant of lactating females have higher feeding demands than non-reproductive females), the one parameter that animals can manipulate in order to reduce the risk of group fragmentation is $r$, the stickiness. This they can do increasing the bonding activities like social grooming that create attraction between individuals \cite{[3],[Dunbar]}. These findings thus reinforce the suggestion that some mechanism to underpin social bonding becomes necessary for animals to be able to live in large stable social groups. 

We have not, at this point, explored the consequences of either the full range of values of $a_{jl}$ or $r$, or of group size. Rather, our aim here has been to establish the principle that uncoupled activity schedules causes fragmentation of groups, and that this can be counteracted by increasing the stickiness, or gravitational pull, between individuals. Finally, it should be noted that in our model we have not attempted to encode day/night preferences, which leads to the animals not aligning their sleep schedule with the day/night cycle. Hence,  as they decouple, males and females end up sleeping at opposite times which would only happen under very particular circumstances. It would be therefore very interesting to extend our model to include further parameters which influence animal cycles, such as the constraints arising from day/night cycles in regions where those are considerable.

\enlargethispage{20pt}


\dataccess{This article has no additional data.}

\aucontribute{The three authors contributed equally to the work.}


\funding{The work of L.P. Schaposnik is partially supported by the Simons Foundation as well as by  the grants NSF DMS FRG 2152107 and NSF CAREER Award DMS 1749013.}

\ack{The  authors  are thankful to MIT PRIMES-USA for the opportunity to conduct this research together.  The material presented here is partially based upon work supported by the National Science Foundation under Grant No. DMS-1928930 while LPS was in residence at the Mathematical Sciences Research Institute in Berkeley, California, during the Fall 2022 semester “Analytic and Geometric Aspects of Gauge Theory” co-organized by Laura Fredrickson (University of Oregon), Rafe Mazzeo (Stanford University), Tomasz Mrowka (Massachusetts Institute of Technology), Laura Schaposnik (University of Illinois at Chicago), and Thomas Walpuski (Humboldt-Universit\"at).}



\end{document}